\documentclass[]{aa520}
\usepackage{graphicx,txfonts,amssymb,natbib}
\authorrunning{Puchwein et al.}
\titlerunning{The impact of gas physics on strong cluster lensing}

\begin{document}

\title{The impact of gas physics on strong cluster lensing}
\author{Ewald Puchwein\inst{1}, Matthias Bartelmann\inst{1},
    Klaus Dolag\inst{2,3}, Massimo Meneghetti\inst{1}
    \institute{$^1$ Zentrum f\"ur Astronomie der Universit\"at
    Heidelberg, ITA, Albert-\"Uberle-Str.~2, 69120 Heidelberg,
    Germany\\
    $^2$ Dipartimento di Astronomia, Universit\`a di Padova, Vicolo
    dell'Osservatorio 2, 35120 Padova, Italy\\
    $^3$ Max-Planck-Institut f\"ur Astrophysik, P.O. Box 1523, 85740
    Garching, Germany}}
\date{\emph{Astronomy \& Astrophysics, submitted}}

\newcommand{\abstext}
 {Previous studies of strong gravitational lensing by galaxy clusters
  have neglected the potential impact of the intracluster gas. Here, we
  compare simulations of strong cluster lensing including gas physics
  at increasing levels of complexity, i.e.~with adiabatic, cooling,
  star-forming, feedback-receiving, and thermally conducting gas, along with
  different implementations of the artificial viscosity in the SPH
  simulations. Each cluster was simulated starting from the same
  initial conditions so as to allow direct comparison of the simulated
  clusters.

  We compare the clusters' shapes, dynamics, and density profiles and then
  study their strong-lensing cross sections computed by means of
  ray-tracing simulations. With the common viscosity implementation,
  adiabatic gas has little effect on strong cluster lensing,
  while lower viscosity allows stronger turbulence, thus higher non-thermal
  pressure and a generally broader gas distribution, which tends to lower lensing cross
  sections. Conversely, cooling and star formation steepen the core
  density profiles and can thus increase the strong-lensing efficiency
  considerably.}

\abstract{\abstext}

\maketitle

\section{Introduction}

How is strong lensing by galaxy clusters affected by the intra-cluster
gas? So far, only the dark-matter distribution was considered for
lensing, and it was ignored that on average approximately 10\% to 15\%
of the cluster matter is contributed by the baryonic gas. This
apparently small contribution may give rise to effects which can
potentially change strong lensing by clusters considerably. Having
seen that the central matter concentration in clusters, their
asymmetries, and their substructures are all very important for the lensing
effects they may create, and considering the non-linear properties of
strong lensing \citep[e.g.][]{BA94.1,BA95.1,ME03.1,TO04.1}, the gas
physics may in fact have substantial consequences.

First, the finite pressure of the hot gas prevents it from forming
density cusps. Thus, the gas distribution has to develop a more or
less extended flat core which reduces the central density compared to
a cluster composed only of dissipation-less dark matter. Second, the
isotropic gas pressure tends to reduce asymmetries in the matter
distribution and to reduce the matter concentration in cluster
substructures. This, in turn, reduces the gravitational tidal field
(the shear) of the cluster mass distribution. Both effects tend to act
against strong cluster lensing and are, therefore, suspected to reduce
the strong-lensing efficiency.

On the other hand, once gas is dense enough to cool on sufficiently
short time scales, it can flow towards the centre and accumulate
there. If it forms stars instead, those act as another component of
dissipation-less matter and may build up matter densities in cluster
cores which exceed those expected in dark-matter clusters. It is not
clear from the start which of these effects will dominate and whether
their competition always leads to the same result. As clusters
form through subsequent mergers, their gas content will also modify
the dynamics of the sub-halo accretion, shrinking the orbits of
infalling sublumps and transporting part of their orbital angular
momentum to the main body of the forming cluster due to friction and
dissipation in the gas. Thus, the gas physics will change not only the
density profile, the level of substructure, and the degree of asymmetry
of clusters, but also modify their internal dynamics following merger
events.

We use gas-dynamical cluster simulations here to investigate the
impact of gas physics on strong lensing by clusters. Most galaxy
clusters were simulated five times; with adiabatic gas, with adiabatic
gas and a novel implementation of artificial viscosity, with cooling
gas which forms stars, with cooling gas which forms stars and conducts
heat, and with only dark matter for reference. In addition, we extracted
cluster-sized halos from a super-cluster simulation which was performed
with only dark matter and including adiabatic gas. All simulations for
a single cluster start from the same initial conditions at early times
so as to allow the cluster simulations to be directly compared. This
cluster sample is described in Sect.~2. Its strong-lensing effects are
studied in Sect.~3, and the results are summarised and discussed in
Sect.~4.

\section{The cluster sample and its physical properties}
\label{sec:cluster_sample}

Our simulations were carried out with {\small GADGET-2}, a new version
of the parallel TreeSPH simulation code {\small GADGET}
\citep{SP01.1}. It uses an entropy-conserving formulation of SPH
\citep{SP02.1}, and upon request provides radiative cooling, heating
by a UV background, and a treatment of star formation and feedback
processes. The latter is based on a sub-resolution model for the
multi-phase structure of the interstellar medium \citep{SP03.1}. For
some of the cluster simulations we also used the new method for
describing heat conduction in SPH, which is both stable and manifestly
conserves thermal energy even when individual and adaptive time-steps
are used \citep{JU04.1}. This implementation assumes an isotropic
effective conductivity parameterised as a fixed fraction of the
Spitzer rate. It also accounts for saturation, which can become
relevant in low-density gas. 


The usual parameterisation of the artificial
viscosity \citep{MO83.1,BAL95.1}
in SPH for an interaction of two particles
$a$ and $b$ includes terms to account for shear and bulk
viscosity.  It is switched on only if particles are approaching; and
for usual cosmological SPH simulations, it can be written as
\begin{equation}
\Pi_{ab}=\frac{-\alpha c_{ab}\mu_{ab}+\beta\mu_{ab}^2}{\rho_{ab}}f_{ab}
\label{eqn:visc}
\end{equation}
for $\vec{r}_{ab}.\vec{v}_{ab}\le 0$ and $\Pi_{ab} = 0$ otherwise, using
\begin{equation}
\mu_{ab}=\frac{h_{ab}\vec{v}_{ab}\cdot\vec{r}_{ab}
}{\vec{r}_{ab}^2+\eta^2}.
\end{equation}
Here $c_{ab}$, $\;\rho_{ab}$, and $h_{ab}$ are the arithmetic means of the sound speed, the density, and the smoothing length, respectively. $\vec{r}_{ab}=\vec{r}_a-\vec{r}_b$ and
$\vec{v}_{ab}=\vec{v}_a-\vec{v}_b$ are the interparticle distance and
the relative velocity. $f_{ab}$ is the mean between particles $a$ and $b$ of the viscosity limiting factor
\begin{equation}
f_i=\frac{|\langle\vec{\nabla}\cdot\vec{v}\rangle_i|}{|\langle\vec{\nabla}\cdot\vec{v}\rangle_i| +
|\langle\vec{\nabla}\times\vec{v}\rangle_i|+\sigma_i}
\end{equation}
which avoids spurious angular momentum and vorticity
transfer in galactic disks, as suggested by
\cite{STE96.1}.
The usual choice for the parameters is $\alpha=0.75$, $\beta=2\alpha$, $\eta=0.01
h_{ab}$, and $\sigma_i=0.0001 c_i/h_i$.

Some simulations were carried out using a
modified artificial viscosity scheme suggested by \cite{MO97.1}, where
every particle evolves its own viscosity
parameter $\alpha_i$, which changes with time according to
\begin{equation}
\frac{d\alpha_i}{dt}=-\frac{\alpha_i-\alpha_{min}}{\tau}+S_i.
\end{equation}
This causes $\alpha_i$ to decay to a minimum value
$\alpha_{min}=0.01$ with e-folding time $\tau$, which we adjust so that
$\alpha_i$ decays over two smoothing lengths after the shock. 
The source term $S_i$, which causes $\alpha_i$ to grow as particles approach a
shock, was assumed to be
\begin{equation}
S_i = S^* f_i \mathrm{max}(0,-|\langle\vec{\nabla}\cdot\vec{v_i}\rangle_i|).
\end{equation}
We choose $S^*=0.7$. Further details on this implementation and its consequences for the generation of turbulence within the intra-cluster medium will be described by Dolag et al. 2005
(in prep).

\subsection{The cluster sample}

We used simulations of four massive galaxy clusters spanning a
mass-range between $1.3\times 10^{15}\:h^{-1}M_\odot$ and $2.3\times
10^{15}\:h^{-1}M_\odot$. The cluster regions were extracted from a
dissipation-less (dark-matter-only) simulation with a box-size of
$479\,h^{-1}\,{\rm Mpc}$ of a flat $\Lambda$CDM model with
$\Omega_0=0.3$, $h=0.7$, $\sigma_8=0.9$, and $\Omega_{\rm b}=0.04$ (see \citealt{YO01.1}).

Using the ``Zoomed Initial Conditions'' (ZIC) technique
\citep{TO97.1}, they were re-simulated with higher mass and force
resolution by populating their Lagrangian volumes in the initial
domain with more particles, appropriately adding additional
small-scale power. The initial particle distributions (before
displacement) are of glass type \citep{WH96.1}.

Gas was introduced into the high-resolution region by splitting each
parent particle into a gas and a dark-matter particle. Thereby, the
gas and the dark-matter particles were displaced by half the original
mean inter-particle distance, such that the centre-of-mass and the
momentum were conserved. The final mass resolution of these
simulations was $m_{\rm DM}=1.13\times 10^9\:h^{-1}M_\odot$ and
$m_{\rm gas}=1.7\times 10^8\:h^{-1}M_\odot$ for dark matter and gas
particles within the high-resolution region, respectively. Thus, the
clusters were resolved with between $2\times10^6$ and $4\times10^6$
particles, depending on their final mass. In addition, we used the
three most massive haloes from a re-simulation of a super-cluster
region, originating from the same cosmological parent simulation,
performed with the same resolution. These three haloes range in mass
between $0.8\times 10^{15}\:h^{-1}M_\odot$ and $1.5\times 10^{15}\:h^{-1}M_\odot$.

For all simulations, the gravitational softening length was kept fixed
at $\epsilon=30.0\,h^{-1}\,\mathrm{kpc}$ comoving
(Plummer-equivalent), and was switched to a physical softening length of
$\epsilon=5.0\,h^{-1}\,\mathrm{kpc}$ at $1+z=6$.

Selection of the initial region was done with an iterative process
involving several low-resolution, dissipation-less re-simulations to
optimise the simulated volume. The iterative cleaning process ensures
that all these haloes are free of contaminating boundary effects up
to at least 3 to 5 times the virial radius. With this usable volume
being relatively large, the simulation also accurately resolves the
clusters' vicinity and thus includes the effects of all the filaments
connected to the cluster.

We used five types of simulations of this galaxy-cluster set. They
comprise simulations with only dark matter (DM), simulations following the
adiabatic evolution of gas but ignoring radiative cooling
(GAS), and simulations including radiative cooling, heating by a UV
background, and a treatment of the star formation and feedback processes
(CSF). The feedback scheme was calibrated to produce a wind
velocity of $\approx350\,\mathrm{km\,s^{-1}}$. Another simulation type
we used additionally includes thermal conduction, for which a fixed
rate of $\kappa=1/3$ times the Spitzer rate was chosen
(CSFC). This choice for $\kappa$ is appropriate in the presence
of magnetised domains with randomly oriented $B$-fields
\citep[e.g.][]{SA88.1} or for a chaotically tangled magnetic field
\citep{NA01.1}. Finally, we used one type of adiabatic gas simulations
in which we applied an implementation of the artificial viscosity, which
damps the build-up of viscosity in the time domain and thus reduces it
considerably where it is not needed (GAS\_NV). The lower
artificial viscosity and the absence of a limiting physical viscosity
allow strong turbulence to build up in the centres of the simulated
clusters. Its contribution to the pressure leads to a significant
density decrease in the cluster cores. For more details, see Dolag et
al. 2005 (in prep.).

\subsection{Halo shape and particle angular momentum}
\label{sec:halo_shape}

\begin{figure}[ht]
\scalebox{0.7}
{
\begin{picture}(0,0)%
\includegraphics{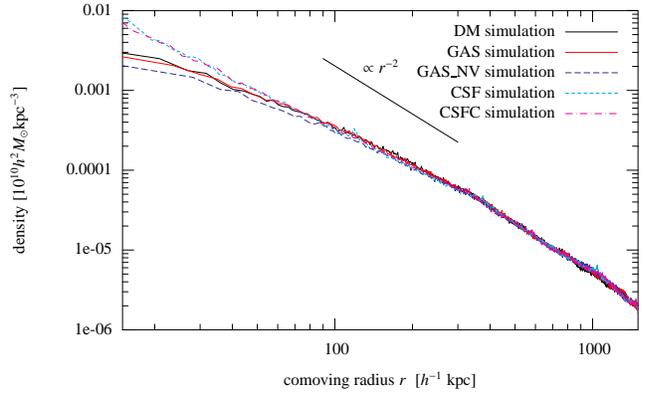}%
\end{picture}%
\setlength{\unitlength}{0.0200bp}%
\begin{picture}(18000,10800)(0,0)%
\put(3025,1650){\makebox(0,0)[r]{\strut{} 1e-06}}%
\put(3025,3800){\makebox(0,0)[r]{\strut{} 1e-05}}%
\put(3025,5950){\makebox(0,0)[r]{\strut{} 0.0001}}%
\put(3025,8100){\makebox(0,0)[r]{\strut{} 0.001}}%
\put(3025,10250){\makebox(0,0)[r]{\strut{} 0.01}}%
\put(9016,1100){\makebox(0,0){\strut{} 100}}%
\put(15953,1100){\makebox(0,0){\strut{} 1000}}%
\put(550,5950){\rotatebox{90}{\makebox(0,0){\strut{density} $[10^{10}h^2M_\odot$kpc$^{-3}]$}}}%
\put(10237,275){\makebox(0,0){\strut{comoving radius} $r\;\;[h^{-1}$ kpc]}}%
\put(10750,8750){\makebox(0,0)[r]{$\propto r^{-2}$}}%
\put(14950,9675){\makebox(0,0)[r]{\strut{}DM simulation}}%
\put(14950,9125){\makebox(0,0)[r]{\strut{}GAS simulation}}%
\put(14950,8575){\makebox(0,0)[r]{\strut{}GAS\_NV simulation}}%
\put(14950,8025){\makebox(0,0)[r]{\strut{}CSF simulation}}%
\put(14950,7475){\makebox(0,0)[r]{\strut{}CSFC simulation}}%
\end{picture}
}
\caption{Profiles of the total density of cluster \emph{g1} at redshift $z=0.2975$
  for different gas-physical models.}
\label{fig:g1_density_pro}
\end{figure}

Before discussing the lensing properties of our simulated galaxy
clusters, we compare their shapes, density profiles, and angular
momentum distributions for the five different physical gas models used
in the simulations. Figure~\ref{fig:g1_density_pro} shows typical
profiles of the total density (dark matter and baryons) for the different types of gas physics. The most
obvious difference is the steeper inner slope in the simulations with
cooling and star formation. Although we use a state-of-the-art
implementation of cooling, feedback, and star formation \citep{SP03.2},
we should point out that it is not entirely clear how realistic the
profiles for these simulations are close to the cluster centre, as the
central cD galaxy will contribute substantially to the core density
profiles \citep{LEW00.1,YO02.1}. The core density of stars in the simulations is larger than
observed (the simulated clusters seem to suffer to some degree from
over-cooling), but some authors (see \citealt{KR05.1}) argue that part
of the discrepancy may be due to stellar mass missed in the
observations.

Despite the isotropic thermal gas pressure, the density profile in the
GAS model is not significantly shallower than in the DM simulation.
This can be understood from the fact that gas particles can reduce their angular
momentum by collisions (see below), which lets them sink towards the cluster centre
more easily. In the GAS\_NV simulation, the additional pressure support due to strong turbulence
allowed by the lower viscosity reduces the gas density in the inner
region of the simulated cluster (see Dolag et al. 2005, in prep., for
more detail). However, the impact of the turbulence on the density
profiles of real clusters requires further investigation because there
the physical viscosity, which is not yet included in the simulations,
may or may not limit the amount of turbulence to smaller values.

\begin{figure}[ht]
\scalebox{0.7}
{
\begin{picture}(0,0)%
\includegraphics{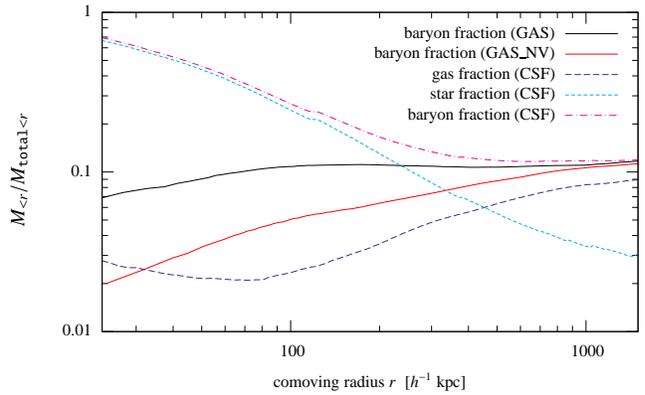}%
\end{picture}%
\begingroup
\setlength{\unitlength}{0.0200bp}%
\begin{picture}(18000,10800)(0,0)%
\put(2475,1650){\makebox(0,0)[r]{\strut{} 0.01}}%
\put(2475,5950){\makebox(0,0)[r]{\strut{} 0.1}}%
\put(2475,10250){\makebox(0,0)[r]{\strut{} 1}}%
\put(7825,1100){\makebox(0,0){\strut{} 100}}%
\put(15775,1100){\makebox(0,0){\strut{} 1000}}%
\put(550,5950){\rotatebox{90}{\makebox(0,0){\strut{} \large $M_{<r} / M_{\mathtt{total} < r}$}}}%
\put(9962,275){\makebox(0,0){\strut{}comoving radius $r\;\;[h^{-1}$ kpc]}}%
\put(14950,9675){\makebox(0,0)[r]{\strut{}baryon fraction (GAS)}}%
\put(14950,9125){\makebox(0,0)[r]{\strut{}baryon fraction (GAS\_NV)}}%
\put(14950,8575){\makebox(0,0)[r]{\strut{}gas fraction (CSF)}}%
\put(14950,8025){\makebox(0,0)[r]{\strut{}star fraction (CSF)}}%
\put(14950,7475){\makebox(0,0)[r]{\strut{}baryon fraction (CSF)}}%
\end{picture}%
\endgroup
}
\caption{Baryonic mass fraction $M_{<r} / M_{\mathtt{total} < r}$ in the GAS and GAS\_NV simulations,
  and mass fractions of gas, stars, and all baryons in the CSF
  simulation, of cluster \emph{g1} at redshift $z=0.2975$ as a
  function of distance $r$ from the cluster centre.}
\label{fig:g1_bar_frac}
\end{figure}

The effects discussed here can also be seen in
Fig.~\ref{fig:g1_bar_frac}, which shows the baryonic mass fractions
for the GAS and GAS\_NV simulations, and the fractions of gas, stars,
and the total baryon fraction for the CSF simulation of cluster
\emph{g1} as a function of the radius of the sphere around the cluster centre in which it was computed.
The baryon fractions of the GAS and CSF simulations are also in good
agreement with the results obtained by \cite{KR05.1}.

\begin{figure}[ht]
\scalebox{0.7}
{
\begin{picture}(0,0)%
\includegraphics{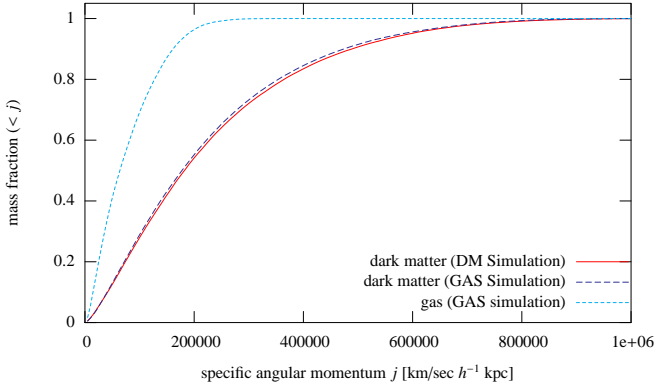}%
\end{picture}%
\setlength{\unitlength}{0.0200bp}%
\begin{picture}(18000,10800)(0,0)%
\put(2200,1650){\makebox(0,0)[r]{\strut{} 0}}%
\put(2200,3288){\makebox(0,0)[r]{\strut{} 0.2}}%
\put(2200,4926){\makebox(0,0)[r]{\strut{} 0.4}}%
\put(2200,6564){\makebox(0,0)[r]{\strut{} 0.6}}%
\put(2200,8202){\makebox(0,0)[r]{\strut{} 0.8}}%
\put(2200,9840){\makebox(0,0)[r]{\strut{} 1}}%
\put(2475,1100){\makebox(0,0){\strut{} 0}}%
\put(5415,1100){\makebox(0,0){\strut{} 200000}}%
\put(8355,1100){\makebox(0,0){\strut{} 400000}}%
\put(11295,1100){\makebox(0,0){\strut{} 600000}}%
\put(14235,1100){\makebox(0,0){\strut{} 800000}}%
\put(17175,1100){\makebox(0,0){\strut{} 1e+06}}%
\put(550,5950){\rotatebox{90}{\makebox(0,0){\strut{mass fraction} $(<j)$}}}%
\put(9825,275){\makebox(0,0){\strut{specific angular momentum} $j$ [km/sec $h^{-1}$ kpc]}}%
\put(15430,3288){\makebox(0,0)[r]{\strut{}dark matter (DM Simulation)}}%
\put(15430,2738){\makebox(0,0)[r]{\strut{}dark matter (GAS Simulation)}}%
\put(15430,2188){\makebox(0,0)[r]{\strut{}gas (GAS simulation)}}%
\end{picture}%
}
\caption{Cumulative distributions of the specific angular momentum of
  the particles contained in the central region of cluster \emph{g1} at $z=0.2975$, 
  within a radius of $r=250\,h^{-1}\,\mathrm{kpc}$.}
\label{fig:g1_angm_pro}
\end{figure}

Figure~\ref{fig:g1_angm_pro} displays the mass fraction of the
particles with specific angular momentum $|\vec L|/m<j$ against the
threshold $j$ for the cluster \emph{g1} at redshift $z=0.2975$. Here,
$\vec L$ and $m$ are the angular momentum and the mass of the
cluster particles, respectively. This quantity is plotted for the
particles of the simulations with only dark matter and for the gas
and dark matter particles of the simulation including adiabatic
gas. The angular-momentum profiles of the GAS\_NV, CSF, and CSFC
simulations, which are not plotted for clarity, are qualitatively
similar to the GAS case. Only particles with a distance smaller than
$250\,h^{-1}\,\mathrm{kpc}$ from the cluster centre were included.

We see from this figure that the profiles of the dark-matter particles
in the two different simulations are almost identical. However, the
gas particles typically have a significantly lower specific angular
momentum, and the same behaviour is also found in the simulations with
cooling and star formation and in the simulations with the new model
for artificial viscosity. When studying the time evolution of these
profiles, we found that the specific angular momenta of both dark
matter and gas particles are boosted towards higher values during
mergers. Afterwards, the angular-momentum profile of the dark matter
is almost conserved, while the gas relaxes and the particles lose
specific angular momentum in collisions. This happens because the
cluster halos lack a well-defined rotation axis and the orbital planes
of gas particles have essentially random orientations. Collisions tend to
average out differences in orbit orientation and thereby reduce the specific angular
momentum of gas particles. Therefore, the difference between the specific angular momentum
of dark matter and gas increases with each merger. We thus find somewhat larger
deviations in large halos, which have on average experienced more
mergers. Reducing their specific angular momenta allows the gas
particles to sink more easily towards the cluster centre. A similar
behaviour of the specific angular momenta of gas and dark matter was
found by \cite{NA91.1} and \cite{NA94.1} in galaxy-sized halos.

By comparing the monopole to the quadrupole moment in circular shells
around the cluster centre, we recover the result obtained by \cite{KA04.1},
who found that halos are more spherical in the simulations with cooling,
feedback, and star formation than in the dissipation-less and adiabatic gas
simulations. For example at the scale radius we find 15 percent smaller ratios of
quadrupole to monopole moments in projections of simulations with cooling and star formation.
We also find more substructure in the form of small clumps in the CSF and CSFC models.

\section{Strong lensing cross sections}

\subsection{Numerical methods}

The centre of each cluster is found by using the halo-finder algorithm
discussed in \cite{TO04.2}. It estimates the dark-matter density at
the position of each dark-matter particle by determining the distance
to the tenth-closest neighbour $d_{10}$ and by also assuming that the density
at the particle position is proportional to $d_{10}^{-3}$. Starting
there, the virial sphere of the particle distribution is found, in
which the gravitational potential is determined. The halo centre is
then taken to be at the potential minimum.

For studying the lensing properties of each cluster, we chose a
sphere of comoving radius $3\,h^{-1}\,\mathrm{Mpc}$ around the cluster
centre and projected all cluster particles inside this region onto an
equidistant grid with a resolution of $4096\times4096$ cells. For the
projection we used the same cubic spline function which is used in the
GADGET code as the SPH smoothing kernel \citep{SP01.1}. We
projected by calculating the overlap of the projected spline function
and the square representing the pixels of the grid.

\begin{figure}[ht]
  \includegraphics[width=\hsize]{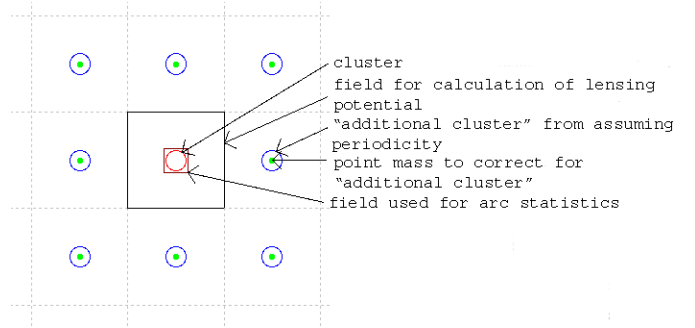}
\caption{Illustration of our method for correcting the error caused by
         the implicit assumption of a periodic input function in fast-Fourier techniques.}
\label{fig:periodicity_correction}
\end{figure}

From the projected mass map, we calculated the convergence $\kappa$ and
its Fourier transform $\hat{\kappa}$, which is related to the Fourier
transform of the lensing potential $\hat{\psi}$ by
\begin{equation}
  \hat{\psi}=-\frac{2}{k^2}\hat{\kappa}\;.
\label{eq:psihat}
\end{equation}
Employing fast-Fourier methods for deriving the lensing potential
according to this equation is, however, problematic. Discrete Fourier
transform algorithms assume that the function to be transformed is
periodic on its support, which is not the case for isolated and finite
cluster convergence fields. Equation~(\ref{eq:psihat}), therefore, does
not yield the lensing potential of a single cluster, but that of an
infinite, two-dimensional \textit{array of clusters} (see
Fig.~\ref{fig:periodicity_correction}), in which the original cluster
is repeated on a grid whose periodicity is set by the side length of a
single cluster field ($6\,h^{-1}\,\mathrm{Mpc}$ comoving in
our case).

Thus, a sufficiently accurate result is only achieved close to the
centre of the cluster field. For reducing the error, we could surround
the cluster by an even larger zero-padded field. However, doing so
without losing resolution substantially increases the demands on
computer memory and slows down the computation. We have therefore
chosen to use a new method to correct for these errors. We place a point mass
that concentrates the total projected mass of the cluster
at the cluster's centre-of-mass. We then
calculate the lensing potential for an array of these point masses with the
Fourier method mentioned above. Next, we analytically subtract the
lensing potential of the single point mass located at the cluster's
centre-of-mass so as to obtain the lensing potential of an array of
point masses with one mass missing at the position of the original
cluster. Subtracting this from the potential obtained for the
\textit{array of clusters}, we correct for the \textit{additional
clusters} implicitly produced by the fast-Fourier algorithms, which
assume a periodic convergence array. The remaining error comes only
from the higher multipole moments of these \textit{additional
clusters} and can be neglected in the central sixteenth
($1024\times1024$ points) of the grid which we use for doing arc
statistics. From this corrected lensing potential, we calculate the
deflection angle.

For finding the images of a number of sources large enough for
statistical analysis, we follow the method introduced by
\cite{MI93.1,MI93.2} and adapted to non-analytic models by
\cite{BA94.1} and \cite{BA95.1}. A previous version of our algorithm is also discussed in some detail
in \cite{ME00.1}. It places elliptical sources with an equivalent
radius of 0.5~arc seconds on an adaptive grid in the source plane, which is fixed at
redshift 1.5 (the lens redshift is taken to be the redshift of the
simulation snapshot), such that there is an increasing density of
sources close to the caustics. A statistical weight is assigned to each source,
which is given by the area represented by the source.

\begin{figure}[ht]
  \includegraphics[width=\hsize]{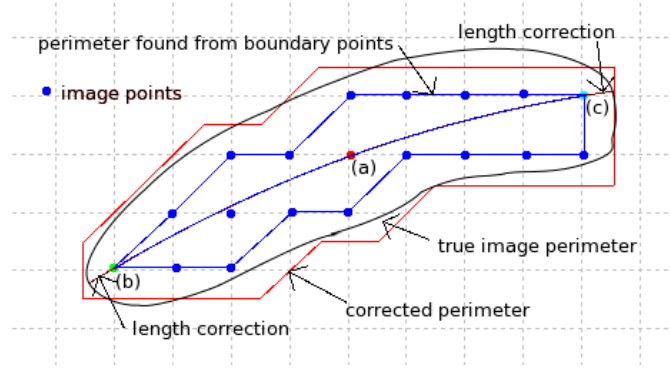}
\caption{Correction of image length and perimeter.}
\label{fig:parameter_correction}
\end{figure}

Next, the deflection angles are used to trace light rays backwards and
map each grid point from the lens plane to the source plane. The
images of a source are found by checking which grid points, when
mapped back to the source plane, are enclosed by the ellipse
corresponding to the source considered. To determine the image
properties (e.g.~length $L$, width $W$, and curvature radius $R$), we
find the image point (a) which, when mapped to the source plane, falls
closest to the source centre. We then find the image point (b) which
is the farthest from (a), and the image point (c), which is the
farthest from (b). The method is illustrated in Fig.~\ref{fig:parameter_correction}.

We could fit a circle through these three points and use the arc
length from (b) to (c) as the length of the image. We would then
determine the image perimeter by walking along the ordered boundary
points and summing up their mutual distances. But since grid cells in
the lens plane are only classified as belonging to the image if their centres fall within the source,
the boundary points of the image (including
(b) and (c)) will on average be about half a grid constant further
inside the image than the true perimeter. Thus, we would
systematically underestimate the length of the image by roughly one
grid constant. We would also underestimate the perimeter of the
image. We correct for this by adding one grid constant to the arc
length from (b) to (c) to obtain the image length, and four grid
constants to the sum of the distances of the boundary points to find
the image perimeter. This is also shown in
Fig.~\ref{fig:parameter_correction}. We found that these new corrections further reduce the weak dependence of
the lensing cross section on the resolution of the grid used for its
computation. 

The image area is calculated directly from the number of
image points. As discussed in \cite{ME00.1}, we search a simple
geometric figure (ellipse, circle, rectangle or ring) with equal area
and length to determine the image width $W$, which is approximated by
the minor axis of the ellipse, the diameter of the circle, the smaller
side of the rectangle or the width of the ring, respectively. We
choose the type of the figure by comparing its circumference to the
perimeter of the image, which we find using the method discussed
above.

We finally determine the lensing cross section $\sigma_{7.5}$ by
summing up the statistical weights of the sources having images with a
length-to-width ratio $L/W\ge7.5$ and calculating the comoving area
corresponding to it. If there is more than one such image for a
source, we multiply the statistical weight of this source by the
number of these images.

\subsection{Results}

We used the method discussed above to find the strong-lensing cross
sections of the clusters \emph{g1}, \emph{g8}, \emph{g51}, \emph{g72},
and of the three largest halos in the super-cluster simulation
\emph{g696}. We did this for 43 simulation snapshots with redshifts
ranging between 0.1 and 1.05, and for the five different physical gas
models discussed in Sect.~\ref{sec:cluster_sample}, except for
\emph{g696}, for which only DM and GAS simulations are available. For
each halo, we used three different projections, namely along the $x$,
$y$ and $z$ axes of the simulation boxes in which the clusters are
randomly oriented.

\begin{figure}[ht]
\scalebox{0.7}
{
\begin{picture}(0,0)%
\includegraphics{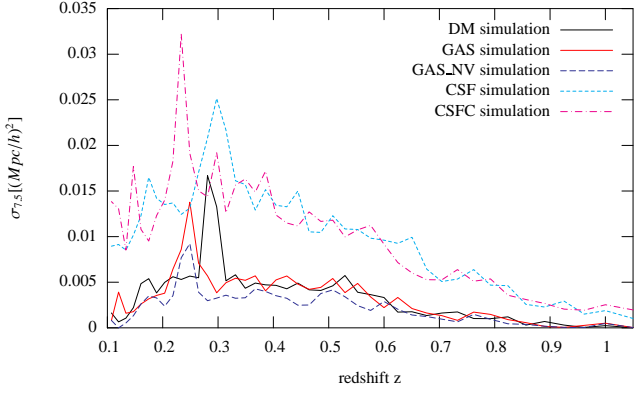}%
\end{picture}%
\setlength{\unitlength}{0.0200bp}%
\begin{picture}(18000,10800)(0,0)%
\put(2750,1650){\makebox(0,0)[r]{\strut{} 0}}%
\put(2750,2879){\makebox(0,0)[r]{\strut{} 0.005}}%
\put(2750,4107){\makebox(0,0)[r]{\strut{} 0.01}}%
\put(2750,5336){\makebox(0,0)[r]{\strut{} 0.015}}%
\put(2750,6564){\makebox(0,0)[r]{\strut{} 0.02}}%
\put(2750,7793){\makebox(0,0)[r]{\strut{} 0.025}}%
\put(2750,9021){\makebox(0,0)[r]{\strut{} 0.03}}%
\put(2750,10250){\makebox(0,0)[r]{\strut{} 0.035}}%
\put(3025,1100){\makebox(0,0){\strut{} 0.1}}%
\put(4514,1100){\makebox(0,0){\strut{} 0.2}}%
\put(6004,1100){\makebox(0,0){\strut{} 0.3}}%
\put(7493,1100){\makebox(0,0){\strut{} 0.4}}%
\put(8983,1100){\makebox(0,0){\strut{} 0.5}}%
\put(10472,1100){\makebox(0,0){\strut{} 0.6}}%
\put(11962,1100){\makebox(0,0){\strut{} 0.7}}%
\put(13451,1100){\makebox(0,0){\strut{} 0.8}}%
\put(14941,1100){\makebox(0,0){\strut{} 0.9}}%
\put(16430,1100){\makebox(0,0){\strut{} 1}}%
\put(550,5950){\rotatebox{90}{\makebox(0,0){\strut{}$\sigma_{7.5} [(Mpc/h)^{2}]$}}}%
\put(10100,275){\makebox(0,0){\strut{}redshift z}}%
\put(14950,9675){\makebox(0,0)[r]{\strut{}DM simulation}}%
\put(14950,9125){\makebox(0,0)[r]{\strut{}GAS simulation}}%
\put(14950,8575){\makebox(0,0)[r]{\strut{}GAS\_NV simulation}}%
\put(14950,8025){\makebox(0,0)[r]{\strut{}CSF simulation}}%
\put(14950,7475){\makebox(0,0)[r]{\strut{}CSFC simulation}}%
\end{picture}%
}
\caption{Strong lensing cross sections for the projection along the
  $y$ axis of the DM, GAS, GAS\_NV, CSF, and CSFC simulations of
  cluster \emph{g51}, for an arc length-to-width ratio of $7.5$ or
  more. The units are comoving $(\mbox{Mpc}/h)^2$.}
\label{fig:g51crossevo}
\end{figure}

\begin{figure}[ht]
\scalebox{0.7}
{
\begin{picture}(0,0)%
\includegraphics{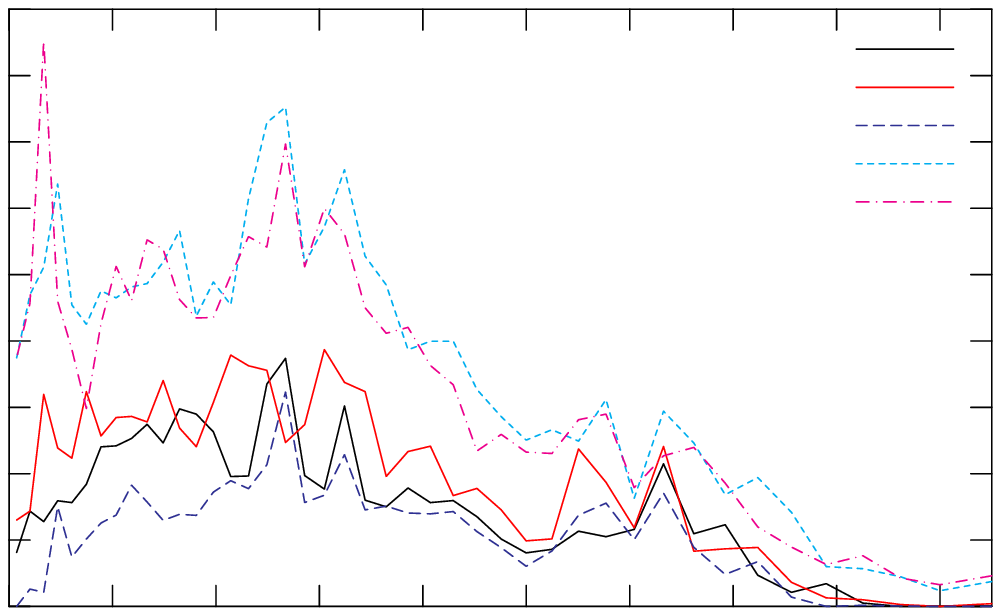}%
\end{picture}%
\setlength{\unitlength}{0.0200bp}%
\begin{picture}(18000,10800)(0,0)%
\put(2750,1650){\makebox(0,0)[r]{\strut{} 0}}%
\put(2750,2606){\makebox(0,0)[r]{\strut{} 0.005}}%
\put(2750,3561){\makebox(0,0)[r]{\strut{} 0.01}}%
\put(2750,4517){\makebox(0,0)[r]{\strut{} 0.015}}%
\put(2750,5472){\makebox(0,0)[r]{\strut{} 0.02}}%
\put(2750,6428){\makebox(0,0)[r]{\strut{} 0.025}}%
\put(2750,7383){\makebox(0,0)[r]{\strut{} 0.03}}%
\put(2750,8339){\makebox(0,0)[r]{\strut{} 0.035}}%
\put(2750,9294){\makebox(0,0)[r]{\strut{} 0.04}}%
\put(2750,10250){\makebox(0,0)[r]{\strut{} 0.045}}%
\put(3025,1100){\makebox(0,0){\strut{} 0.1}}%
\put(4514,1100){\makebox(0,0){\strut{} 0.2}}%
\put(6004,1100){\makebox(0,0){\strut{} 0.3}}%
\put(7493,1100){\makebox(0,0){\strut{} 0.4}}%
\put(8983,1100){\makebox(0,0){\strut{} 0.5}}%
\put(10472,1100){\makebox(0,0){\strut{} 0.6}}%
\put(11962,1100){\makebox(0,0){\strut{} 0.7}}%
\put(13451,1100){\makebox(0,0){\strut{} 0.8}}%
\put(14941,1100){\makebox(0,0){\strut{} 0.9}}%
\put(16430,1100){\makebox(0,0){\strut{} 1}}%
\put(550,5950){\rotatebox{90}{\makebox(0,0){\strut{}$\sigma_{7.5} [(Mpc/h)^{2}]$}}}%
\put(10100,275){\makebox(0,0){\strut{}redshift z}}%
\put(14950,9675){\makebox(0,0)[r]{\strut{}DM simulation}}%
\put(14950,9125){\makebox(0,0)[r]{\strut{}GAS simulation}}%
\put(14950,8575){\makebox(0,0)[r]{\strut{}GAS\_NV simulation}}%
\put(14950,8025){\makebox(0,0)[r]{\strut{}CSF simulation}}%
\put(14950,7475){\makebox(0,0)[r]{\strut{}CSFC simulation}}%
\end{picture}%
}
\caption{Strong lensing cross section for the DM, GAS, GAS\_NV, CSF,
  and CSFC simulation of cluster \emph{g1} and an arc
  length-to-width ratio of $\ge7.5$.}
\label{fig:g1crossevo}
\end{figure}

In Fig.~\ref{fig:g51crossevo}, we plot the cross section as a function
of redshift for one of the projections of \emph{g51}. We can see that
despite the gas pressure, adiabatic gas with the standard artifical
viscosity does not reduce the strong lensing cross section compared to
the dark-matter-only simulation. Using the new artifical viscosity
scheme, however, leads to somewhat smaller cross sections due to the
more extended gas distribution it implies. On the other hand, cooling,
star formation and feedback make the simulated cluster a significantly
more efficient lens. These three properties are typical for most of
the clusters we have studied. There is generally no large difference
between the DM and (adiabatic) GAS models, a somewhat smaller cross
section in the GAS\_NV model, and cross sections larger by a factor of
1.5 to 3 in the CSF and CSFC models. In some cases, however, even the
adiabatic gas with standard artifical viscosity causes an
\emph{increase} in the cross section compared to the simulations
containing only dark matter. This is illustrated in
Fig.~\ref{fig:g1crossevo}.

For determining the impact of cluster ellipticity and substructure on
the strong lensing cross sections, we transform the maps of the
surface mass density of our cluster halos to polar coordinates
(centred on the cluster halo) and average over the polar angle to
obtain maps of an axially symmetrised cluster. Then, we use the same
methods as before to compute strong-lensing cross sections.

\begin{figure}[ht]
\scalebox{0.7}
{
\begin{picture}(0,0)%
\includegraphics{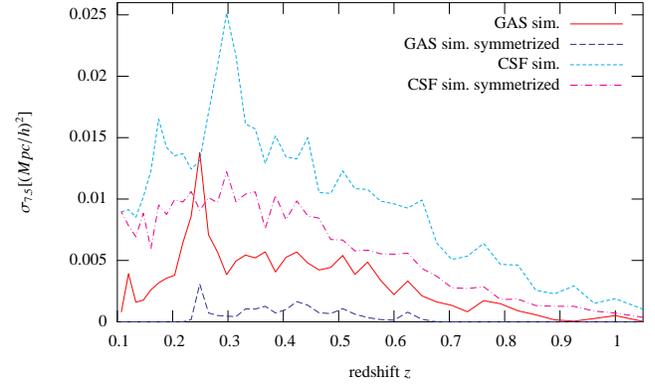}%
\end{picture}%
\setlength{\unitlength}{0.0200bp}%
\begin{picture}(18000,10800)(0,0)%
\put(2750,1650){\makebox(0,0)[r]{\strut{} 0}}%
\put(2750,3304){\makebox(0,0)[r]{\strut{} 0.005}}%
\put(2750,4958){\makebox(0,0)[r]{\strut{} 0.01}}%
\put(2750,6612){\makebox(0,0)[r]{\strut{} 0.015}}%
\put(2750,8265){\makebox(0,0)[r]{\strut{} 0.02}}%
\put(2750,9919){\makebox(0,0)[r]{\strut{} 0.025}}%
\put(3025,1100){\makebox(0,0){\strut{} 0.1}}%
\put(4514,1100){\makebox(0,0){\strut{} 0.2}}%
\put(6004,1100){\makebox(0,0){\strut{} 0.3}}%
\put(7493,1100){\makebox(0,0){\strut{} 0.4}}%
\put(8983,1100){\makebox(0,0){\strut{} 0.5}}%
\put(10472,1100){\makebox(0,0){\strut{} 0.6}}%
\put(11962,1100){\makebox(0,0){\strut{} 0.7}}%
\put(13451,1100){\makebox(0,0){\strut{} 0.8}}%
\put(14941,1100){\makebox(0,0){\strut{} 0.9}}%
\put(16430,1100){\makebox(0,0){\strut{} 1}}%
\put(550,5950){\rotatebox{90}{\makebox(0,0){\strut{}$\sigma_{7.5} [(Mpc/h)^{2}]$}}}%
\put(10100,275){\makebox(0,0){\strut{}redshift $z$}}%
\put(14950,9675){\makebox(0,0)[r]{\strut{}GAS sim.}}%
\put(14950,9125){\makebox(0,0)[r]{\strut{}GAS sim. symmetrized}}%
\put(14950,8575){\makebox(0,0)[r]{\strut{}CSF sim.}}%
\put(14950,8025){\makebox(0,0)[r]{\strut{}CSF sim. symmetrized}}%
\end{picture}%
}
\caption{Strong lensing cross sections for the GAS, CSF, and
  axially-symmetrised GAS and CSF versions of cluster \emph{g51}.}
\label{fig:g51sym}
\end{figure}

Figure~\ref{fig:g51sym} shows the cross sections of the GAS and CSF
versions of \emph{g51} and of its axially-symmetrised GAS and CSF
counterparts. One can clearly see that substructure and ellipticity
significantly increase the strong-lensing cross sections (see also
\citealt{BA95.1} and \citealt{ME03.1}). Note also that the increase of the
cross section in the CSF simulation, compared to the GAS simulation,
is almost the same for the original cluster and its axially-symmetric
variant. This can be interpreted such that this increase in the cross
section in the cooling and star formation simulations is caused mainly
by the steeper density profile and not by any changes of the
ellipticity or the substructure.

There is also no qualitative difference between the lensing properties
of the simulated clusters and the largest halos of the simulated super
cluster region.

\begin{figure}[ht]
\scalebox{1.1}
{
\begin{picture}(0,0)(0,0)%
\includegraphics{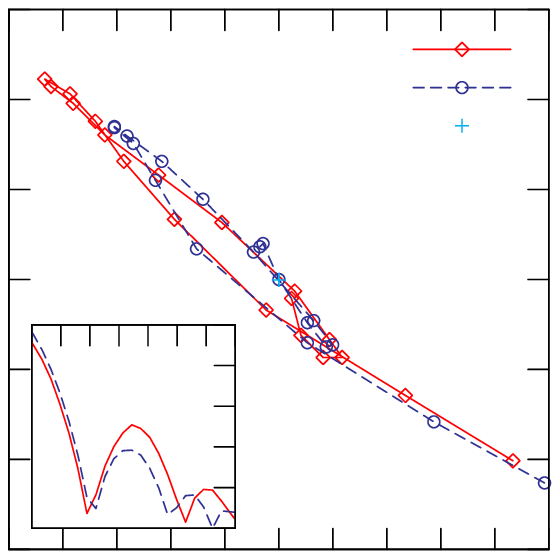}%
\end{picture}%
\begingroup
\setlength{\unitlength}{0.0200bp}%
\begin{picture}(18000,9700)(0,0)%
\put(2200,1650){\makebox(0,0)[r]{\strut{} 1.5}}%
\put(2200,2946){\makebox(0,0)[r]{\strut{} 2}}%
\put(2200,4242){\makebox(0,0)[r]{\strut{} 2.5}}%
\put(2200,5537){\makebox(0,0)[r]{\strut{} 3}}%
\put(2200,6833){\makebox(0,0)[r]{\strut{} 3.5}}%
\put(2200,8129){\makebox(0,0)[r]{\strut{} 4}}%
\put(2200,9425){\makebox(0,0)[r]{\strut{} 4.5}}%
\put(2475,1100){\makebox(0,0){\strut{} 2.5}}%
\put(3253,1100){\makebox(0,0){\strut{} 2.6}}%
\put(4030,1100){\makebox(0,0){\strut{} 2.7}}%
\put(4808,1100){\makebox(0,0){\strut{} 2.8}}%
\put(5585,1100){\makebox(0,0){\strut{} 2.9}}%
\put(6363,1100){\makebox(0,0){\strut{} 3}}%
\put(7140,1100){\makebox(0,0){\strut{} 3.1}}%
\put(7918,1100){\makebox(0,0){\strut{} 3.2}}%
\put(8695,1100){\makebox(0,0){\strut{} 3.3}}%
\put(9473,1100){\makebox(0,0){\strut{} 3.4}}%
\put(10250,1100){\makebox(0,0){\strut{} 3.5}}%
\put(550,5537){\rotatebox{90}{\makebox(0,0){\strut{} \small $h^{-1}$ Mpc}}}%
\put(6362,275){\makebox(0,0){\strut{}$h^{-1}$ Mpc}}%
\put(8025,8850){\makebox(0,0)[r]{\strut{} \scriptsize subhalo position (DM sim.)}}%
\put(8025,8300){\makebox(0,0)[r]{\strut{} \scriptsize subhalo position (GAS sim.)}}%
\put(8025,7750){\makebox(0,0)[r]{\strut{} \scriptsize main halo position}}%
\put(5800,1954){\makebox(0,0)[l]{\strut{} \tiny 0}}%
\put(5800,3126){\makebox(0,0)[l]{\strut{} \tiny 1}}%
\put(5800,4298){\makebox(0,0)[l]{\strut{} \tiny 2}}%
\put(3224,5150){\makebox(0,0){\strut{} \tiny 7}}%
\put(4061,5150){\makebox(0,0){\strut{} \tiny 8}}%
\put(4898,5150){\makebox(0,0){\strut{} \tiny 9}}%
\put(6200,3200){\rotatebox{90}{\makebox(0,0){\strut{} \tiny radial disctance}}}%
\put(6550,3200){\rotatebox{90}{\makebox(0,0){\strut{} \tiny [$h^{-1}$ Mpc]}}}%
\put(4270,5500){\makebox(0,0){\strut{}\tiny time [Gyr]}}%
\end{picture}%
\endgroup
}
\caption{Position (in the main halo's rest frame and in comoving
  coordinates) and radial disctance from the main halo of the sub-halo whose passage
  produces the peaks in cluster \emph{g72}'s lensing cross section for the DM and GAS case.}
\label{fig:g72merger}
\end{figure}

The different physics in the simulations with only dark matter
particles compared to simulations including gas directs sub-halos
passing close to the main halo during a merger into different
orbits. Compared to the dissipation-less simulation with only dark
matter, the sub-halo loses angular momentum and energy in the
gas-dynamical simulations and is directed into a less elliptical
orbit. It therefore returns earlier for the next passage of the main
halo's centre. This is illustrated in Fig.~\ref{fig:g72merger}, which
shows the position of a sub-halo in the rest frame of the main halo of
the cluster \emph{g72}. Note that the $x$ and $y$ axes are scaled
differently for clarity. The sub-halos in the DM and GAS simulations
initially move approximately synchronously and approach the main halo
with almost the same velocity. Later, however, the different dynamics
makes the GAS sub-halo's orbit substantially less elliptical, hence it
returns earlier for the next core passage.

\begin{figure}[ht]
\scalebox{0.7}
{
\begin{picture}(0,0)%
\includegraphics{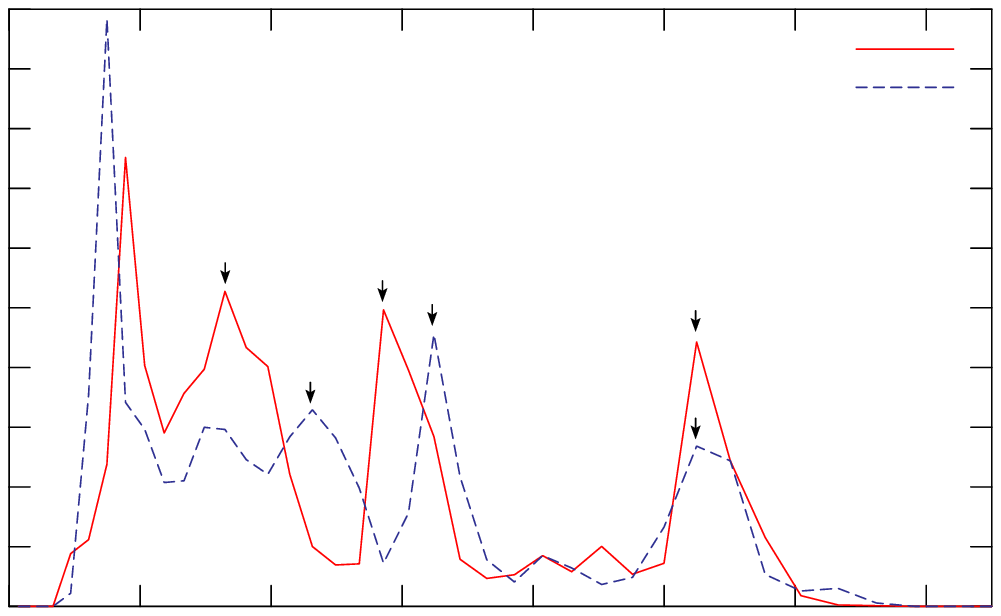}%
\end{picture}%
\setlength{\unitlength}{0.0200bp}%
\begin{picture}(18000,10800)(0,0)%
\put(2750,1650){\makebox(0,0)[r]{\strut{} 0}}%
\put(2750,2510){\makebox(0,0)[r]{\strut{} 0.002}}%
\put(2750,3370){\makebox(0,0)[r]{\strut{} 0.004}}%
\put(2750,4230){\makebox(0,0)[r]{\strut{} 0.006}}%
\put(2750,5090){\makebox(0,0)[r]{\strut{} 0.008}}%
\put(2750,5950){\makebox(0,0)[r]{\strut{} 0.01}}%
\put(2750,6810){\makebox(0,0)[r]{\strut{} 0.012}}%
\put(2750,7670){\makebox(0,0)[r]{\strut{} 0.014}}%
\put(2750,8530){\makebox(0,0)[r]{\strut{} 0.016}}%
\put(2750,9390){\makebox(0,0)[r]{\strut{} 0.018}}%
\put(2750,10250){\makebox(0,0)[r]{\strut{} 0.02}}%
\put(3025,1100){\makebox(0,0){\strut{} 0.1}}%
\put(4912,1100){\makebox(0,0){\strut{} 0.2}}%
\put(6798,1100){\makebox(0,0){\strut{} 0.3}}%
\put(8685,1100){\makebox(0,0){\strut{} 0.4}}%
\put(10572,1100){\makebox(0,0){\strut{} 0.5}}%
\put(12458,1100){\makebox(0,0){\strut{} 0.6}}%
\put(14345,1100){\makebox(0,0){\strut{} 0.7}}%
\put(16232,1100){\makebox(0,0){\strut{} 0.8}}%
\put(550,5950){\rotatebox{90}{\makebox(0,0){\strut{}$\sigma_{7.5} [(Mpc/h)^{2}]$}}}%
\put(10100,275){\makebox(0,0){\strut{}redshift z}}%
\put(14950,9675){\makebox(0,0)[r]{\strut{}DM simulation}}%
\put(14950,9125){\makebox(0,0)[r]{\strut{}GAS simulation}}%
\end{picture}%
}
\caption{Strong-lensing cross section for the DM and GAS simulations
  of cluster \emph{g72} and a length-to-width ratio of $7.5$ or
  more. The arrows mark the three successive passages
  of the same sub-halo for both models.}
\label{fig:g72xasync}
\end{figure}

This effect also has an impact on the strong lensing cross section,
because the cross section of a halo increases when a sub-halo crosses the cluster core
during a merger (see \citealt{TO04.1} and \citealt{ME04.1}). In Fig.~\ref{fig:g72xasync},
the peaks in the cross section corresponding to the three successive
passages, illustrated in Fig.~\ref{fig:g72merger}, are marked by
arrows for both the DM and the GAS simulations. The first peak happens
in both cases at a redshift $z\approx0.62$. The second and the third
peaks, however, occur at significantly earlier times in the GAS
simulation. The peak positions for the GAS\_NV, CSF, and CSFC
simulations are typically very similar to the GAS case. Note that the
different heights of the first peak in the different simulations do
not imply a fundamental difference, because the amplitude of sharp
peaks depends strongly on the exact time when the snapshot was taken
(the time resolution is just a few snapshots for the passage). Thus, a
slight deviation in the timing may result in peaks with apparently
very different heights.

\section{Summary and discussion}

We have compared the density profiles, shapes, and strong lensing
properties of numerically simulated galaxy clusters. The simulations
were performed starting from the same initial conditions, but with
different gas physics. Five different gas-physical models were
employed. They contain:

\begin{itemize}
\item DM model: dark matter only;
\item GAS model: dark matter and adiabatic gas;
\item GAS\_NV model: dark matter, adiabatic gas, and a new
  implementation of the artificial viscosity, reducing the viscosity
  where it is not needed;
\item CSF model: dark matter, cooling gas, a star formation model,
  and feedback;
\item CSFC model: dark matter, cooling gas, a star formation model,
  feedback, and thermal conductivity;
\end{itemize}

Our cluster sample consisted of four simulated galaxy clusters and the
three largest halos of a simulation of a super-cluster region. We used
43 snapshots of these halos between redshifts $1.05$ and $0.1$. For
studying strong lensing, we used the three different projections along
the coordinate axes of the simulation volume, and sources fixed at
redshift $1.5$.

We find significantly steeper inner slopes for the density profiles of
halos simulated with cooling and star formation. On the other hand,
adiabatic gas with a standard artifical viscosity, in spite of its
isotropic thermal pressure, does not make density profiles shallower
compared to the dissipation-less dark-matter simulations. This can be
understood from the fact that gas can reduce its specific angular
momentum by collisions. The orbits of gas particles in a cluster halo
are randomly oriented. Collisions tend to average out these
differences and thereby reduce the specific angular momentum of the
gas, which helps the gas to move towards the cluster centre. This
effect can also be seen in the specific angular-momentum profiles we
present in Fig.~\ref{fig:g1_angm_pro}. However, the additional
pressure caused by strong turbulence in the GAS\_NV simulation
somewhat reduces the density close to the cluster centre.

We then performed ray-tracing simulations for the numerically simulated
galaxy clusters to study their strong lensing properties. We calculated
cross sections for long thin arcs with a length-to-width ratio equal
to or larger than $7.5$. For the simulations with cooling and star
formation, we found significantly larger cross sections. Thermal
conductivity has no big impact on strong lensing. Despite its
pressure, adiabatic gas with standard artifical viscosity does not
reduce the cross section for long thin arcs. In some cases, it even
makes the cluster a more efficient lens. On the other hand, simulating
adiabatic gas with the new scheme for reduced artifical viscosity
leads to somewhat smaller strong lensing cross sections compared to
the DM and GAS runs.

Note that despite our use of a state-of-the-art model for cooling,
star formation, and feedback, the simulated clusters suffer from some
over-cooling. Therefore, the density close to the centre and the
increase of the strong lensing cross section may be over-predicted in
the simulations with cooling and star formation compared to real
clusters. Also, the impact of turbulence on galaxy clusters needs
further investigation. In real clusters, the physical viscosity of the
cluster gas, which is not yet included in simulations, will regulate
turbulence and may lead to a smaller amount of turbulence than in the
GAS\_NV simulations. It will also be interesting to investigate
the effect of this new scheme of artifical viscosity on simulations
including cooling, star formation, and feedback.

To study the impact of ellipticity and substructure on the lensing
properties we transformed maps of the surface mass density of our
cluster halos to polar coordinates and averaged over the angle, so that
we obtained maps of an azimuthally symmetrised halo. We compared the
cross sections of these axially symmetrised clusters to those of the
numerically simulated ones and found that ellipticity and substructure
are important for strong lensing in all of the five gas physical
models. However, their impact on the cross section turns out to be
very similar in the simulations without and with cooling and star
formation. Thus, the larger cross sections we obtained for simulations
with cooling and star formation are mainly caused by the steeper
density profile and not by changes in ellipticity and substructure.

Sub-halos passing close to the main halo lose angular momentum and
energy and are directed into less elliptical orbits in simulations
with gas compared to dark-matter-only simulations. During a merger,
such a sub-halo will return at an earlier time for the next
passage. Thus, mergers proceed faster in simulations with gas. This
can also be seen in the strong lensing cross sections: typically, the
peak corresponding to the first passage of a sub-halo happens at the
same time in the dissipation-less and the gas-dynamical simulations,
but the peaks corresponding to the next passages are shifted to
earlier times in the simulations with gas.

The work presented here clearly shows that cluster gas can have a significant
impact on strong lensing cross sections. However, the importance of the different effects 
(turbulence, cooling, star formation, mergers) is different for each individual cluster and changes during the cluster's
evolution. To infer cluster properties from observations and for studying the importance of these effects in real clusters, it is thus more promising to model observed clusters individually than to do statistical analyses of large cluster samples. For this it will be interesting to combine the lensing information with thermal Sunyaev-Zeldovich effect and X-ray observations. We are planning to simulate X-ray emission and thermal SZ effect for the numerical galaxy clusters used here and to study methods for combining such (mock) observations to model individual clusters and reconstruct their mass distributions.

\acknowledgements{We are deeply indebted to Volker Springel for
providing access to P-GADGET-2 prior to release, and for support in using and modifying it.
The simulations were carried out on the IBM-SP4
machine at the Centro Interuniversitario del Nord-Est per il Calcolo
Elettronico (CINECA, Bologna), with CPU time assigned under an
INAF-CINECA grant, on the IBM-SP3 at the Italian Centre of Excellence
``Science and Applications of Advanced Computational Paradigms'',
Padova, and on the IBM-SP4 machine at the Rechenzentrum der
Max-Planck-Gesellschaft with CPU time assigned to the
Max-Planck-Institut f\"ur Astrophysik. E.~P.~is supported by the
German Science Foundation under grant number
BA~1369/6-1. K.~D.~acknowledges support by a Marie Curie Fellowship of
the European Community program ``Human Potential'' under contract
number MCFI-2001-01227.

\end{document}